\documentstyle[12pt,aaspp4]{article}

\slugcomment{To appear in {\sl The Astrophysical Journal}}

\begin{document}
\title{POWER SPECTRA OF X-RAY BINARIES} 
\author{T. P. Li\altaffilmark{1,2,3} and Y. Muraki\altaffilmark{3}}
  
\altaffiltext{1}{Department of Physics \& Center for Astrophysics, Tsinghua
University,
 Beijing}
\altaffiltext{2} {Institute of High Energy Physics, Chinese Academy of
Sciences, Beijing}
\altaffiltext{3}{Solar-Terrestrial Environment Laboratory, Nagoya University,
Nagoya}

\begin{abstract}
The interpretation of Fourier spectra in the time domain is critically
examined.  Power density spectra defined and calculated in the time domain 
are compared with Fourier spectra in the frequency domain for three different 
types of variability: periodic signals, Markov processes and random shots. 
The power density spectra for a sample of neutron stars and black hole binaries
are analyzed in both the time and the frequency domains. 
For broadband noise, the two kinds of power spectrum in accreting neutron stars
are usually consistent with each other, but the time domain power spectra
for black hole candidates are significantly higher than  
corresponding Fourier spectra in the high frequency range (10--1000 Hz). 
Comparing the two kinds of power density spectra 
may help to probe the intrinsic nature of timing phenomena in compact objects. 
\end{abstract}

\section{INTRODUCTION} 
The Fourier transform is widely used in timing analysis in astronomy. 
Through a discrete Fourier transform, a light curve $x(t_k)$
can be decomposed into sine wave components in the frequency domain
\begin{equation}
a(f_j)=\sum_kx(t_k)e^{-i2\pi f_jk\Delta t}~,
\end{equation}
where $a(f_j),~j=-N/2,\cdots,N/2-1,$ 
is the Fourier amplitude of  sine wave components at frequency $f_j=j\Delta f$
with $\Delta f=1/(N\Delta t)$ and $x(t_k)$ represents the photon counts during
a time interval 
$(t_k, t_k+\Delta t)$, $k=1,2,\cdots,N$,
with $t_k=(k-1)\Delta t$.
The Fourier spectrum can be a powerful technique in the search for periodic
signals, pulsations, and quasi-periodic oscillations in the light curve 
of a X-ray source. The Fourier power density 
\begin{equation}
p_f(f_j)=|a(f_j)|^2
\end{equation}
is used to describe the variability amplitude at different frequency $f_j$.
A peak with power in excess of the noise distribution power in the Fourier
spectrum indicates 
the existence of a periodic component in the light curve. When estimating the
amplitude 
of a periodic signal from the Fourier power density spectrum, one must
assume that the shape of a signal pulse is sinusoidal, which is usually 
not true for variability due to real physical processes. 

Even more care should be taken in interpreting
the Fourier spectrum of an aperiodic process in the time domain. 
It is a prevalent misconception that
the Fourier power spectrum is the only way to express the
distribution of variability amplitude vs. time scale, and that $p_f(f_j)$ is 
a quantity describing the variability amplitude for the time scale
$1/f_j$. The actual relation between the Fourier power and 
the variability process occurring in the time domain is 
given by Parseval's theorem:
\begin{equation}
\frac{1}{N}\sum_{j=-2/N}^{N/2-1}|a(f_j)|^2=\sum_{k=1}^{N}|x(t_k)|^2~,
\end{equation}
or equivalently,
\begin{equation}
\sum_{j}p_f(f_j)\Delta f = \sum_{k}|r(t_k)|^2\Delta t~,
\end{equation}
where $r(t_k)=x(t_k)/\Delta t$ is the counting rate at $t_k$. 
Parseval's theorem states that 
the integral of the Fourier power density over the whole frequency range
is equal to the variability power of the same process in the time domain.
In fact, that is just the reason why the quantity $p_f(f_j)=|a(f_j)|^2$ is
called {\sl power} density.  However, Parseval's theorem says nothing 
about the power density {\it distribution} over the time domain.
The Fourier power at any given frequency within the limited range,
$0 \rightarrow 1/\Delta t$, actually represents
a sum of complex amplitudes from an infinite number of aliased
frequencies, each
contribution of which is reduced by a factor of sinc$(\pi f \Delta t)$,
with the sum of the squares of all the factors being unity.
In contrast, the power at any given frequency in a time-based spectrum
is most sensitive to the {\it rate of change} of the Fourier spectrum
with frequency near the Nyquist frequency ($f~=~1/(2 \Delta t)$) and its
aliases, since the
sinc function given above is changing most rapidly for such
frequencies.
The rms variation vs. time scale of a time series 
may, in fact, differ substantially from its Fourier
spectrum. Even in the simplest case, where the signal 
is purely sinusoidal with a frequency $f$, 
the Fourier spectrum is a $\delta$ function for the continuous Fourier
transform 
or approximately a $\delta$ function for the discrete case. 
Such a function has little power density at frequencies except those near
$f$, 
which, although in this case is a clear representation of what 
physical process is going on, is still not an 
accurate representation of the variability of the amplitude in a time series. 
Only when the time step is set equal to or much greater than the period,
$\Delta t\ge 1/f$, 
can the corresponding light curve (or pulse profile) of a sinusoidal 
signal be completely flattened to make the time variation vanish.
At time scales shorter than $1/f$, however, a variation of intensity still
exists 
in the light curve, and under some circumstances one would wish that 
power over this region should not be zero. In fact, different but
mathematically
equivalent representations with different bases 
or functional coordinates in the frequency domain do exist for certain light
curves.
The Fourier transform with its trigonometric basis is just one of many possible 
transforms between the time and frequency domains, and may not necessarily
represent
the true power distribution of a {\it physical} process in the time domain. 
   
One can also argue that any observable physical process always occurs in the
time domain,
and this is why real variability amplitudes at different time scales 
are useful in understanding a time varying process, in addition to the
understanding
gleaned from the conventional Fourier spectrum.   
An algorithm to calculate the power density spectrum directly in the time
domain 
without using the Fourier transform has been proposed (\cite{lit01}).
We introduce the algorithm and compare the power spectra in the time domain
with the Fourier spectra for different kinds of time process model in \S 1.       
We have studied the power density spectra of a sample of neutron stars
and black hole binaries,
in both the frequency {\it and} the time domains. 
The results are presented in \S 3. 
A discussion of the potential gain in the understanding the intrinsic nature 
of physical processes occurring near compact objects via comparison 
of the two kinds of power density spectrum is given in \S 4.

\section{POWER SPECTRA IN THE TIME DOMAIN}
The initial definition of the variation power in a light curve $x(k)$ is
\begin{equation}
P(\Delta t)=\frac{\mbox{Var}(x)}{(\Delta
t)^2}=\frac{\frac{1}{N}\sum_{k=1}^{N}(x(k)-\bar{x})^2}
{(\Delta t)^2}=\frac{1}{N}\sum_{k=1}^{N}(r(k)-\bar{r})^2,
\end{equation}
where the light curve $x(k),~k=1,\dots,N$, is a counting series obtained from a
time
history of observed photons with a time step $\Delta t$, $r(k)$ is the
corresponding 
rate series,   
and $\bar{x}=\sum_{k=1}^{N}x(k)/N$ and $\bar{r}=\sum_{k=1}^{N}r(k)/N$ are the
average
counts and count rates, respectively.
The unit used for a power $P$ in spectral analysis in astronomy is usually 
rms$^2$, where 'rms' refers to the analysed time series $x$.
In the case of $x$ being a counting series, the power $P$ can be expressed 
in units of counts$^2$/s$^2$. 
$P(\Delta t)$ represents an integral of variability power for the region
of time scale
$\ge \Delta t$.  It is easy to prove that $P(\Delta t_1)\ge P(\Delta t_2)$ 
if $\Delta t_2 > \Delta t_1$.  Thus we can define the power density $p(\Delta
t)$ 
in the time domain as the rate of change of $P(\Delta t)$ with respect to the
time
step $\Delta t$ 
\begin{equation}
p(\Delta t)=\frac{d P(\Delta t)}{d\Delta t},
\end{equation}
in counts$^2$/s$^3$ or rms$^2$/s.
With Eq. (5) and (6), we can calculate the power density spectrum in the time
domain
for a light curve. In practice, the differential calculus in Eq. (6) can be
performed numerically
\begin{equation}
p(\Delta t)=\frac{P(\Delta t_1)-P(\Delta t_2)}{\Delta t_2-\Delta t_1},
\end{equation}
with $P(\Delta t_1)$ and $P(\Delta t_2)$ being the two powers 
at the time scale $\Delta t_1$ and $\Delta t_2$ ($\Delta t_2>\Delta t_1$),
respectively, and
$\Delta t=(\Delta t_1+\Delta t_2)/2$.

To detect the source signal in a power spectrum against a noise background,
 we need to know the noise power $P_{noise}(\Delta t)$ of a time series
consisting only of noise.
For a white noise series where the $x(k)$ follows
 the Poisson distribution, the noise power is
\begin{equation}
P_{noise}(\Delta t)=\frac{\mbox{Var}(x)}{(\Delta t)^2}=\frac{<x>}{(\Delta t)^2}
=\frac{r}{\Delta t},
\end{equation}
where $<x>$ is the expectation value of $x$, but with units of counts$^2$, 
which also the variance of the Poisson variable $x$, and $r$ is the 
expectation value of the counting rate which can be estimated by 
the global average of counting rate of the studied observation.
The noise power density  at $\Delta t=(\Delta t_1+\Delta t_2)/2$ is
\begin{equation}
p_{noise}(\Delta t)=\frac{P_{noise}(\Delta t_1)-P_{noise}(\Delta t_2)}
{\Delta t_2-\Delta t_1}
 =\frac{r}{\Delta t_1 \Delta t_2}. 
\end{equation}
The signal power density can be defined as
\begin{equation}
p_{signal}(\Delta t)=p(\Delta t)-p_{noise}(\Delta t) 
 \end{equation}
and the fractional signal power density as
\begin{equation}
p'_{signal}(\Delta t)=\frac{p_{signal}(\Delta t)}{r^2},
\end{equation}
in units of 1/s or (rms/mean)$^2$/s, where both 'rms' and 'mean' refer 
to the time series.

 To study the signal power density in the time domain over a background 
of noise in an observed photon series, we divide the observation 
into $M$ segments. For each segment $i$,
 the signal power density $p_{i,signal}(\Delta t)$ is
 calculated by Eq.~(10). The average power density of the studied observation
is
$\bar{p}_{s}=\sum_{i=1}^M p_{i,signal}/M$ and its standard deviation is
$\sigma(\bar{p}_{s})=\sqrt{\sum_{i=1}^{M}(p_{i,signal}-\bar{p}_{s})^2/(M(M-1))}
$.
We can use the statistical methods based on
the normal distribution to make statistical inference, e.g. significance test,
on $\bar{p}_s$. At short time steps, the number  $x$
of counts per bin may be too small for it to behave as
 a normally distributed variable. But it is easy
to get a large enough total number ($M$) of segments 
from a certain observation period 
to satisfy the condition for applying the central limit theorem    
as well as for using the normal statistics on the mean $\bar{p}_s$. 

To compare the power spectrum in the time domain with the 
Fourier spectrum for the same process, we study three 
different kinds of time series.

(1) {\sl Periodic signal}

We use Eq.(5) and (6) to calculate the power density spectrum in the time
domain
for a sinusoidal process with a period of 0.5 s. In Figure 1, a piece of the
counting 
rate curve of the studied process is shown in the top panel and the  power 
density distribution of time scale of the sinusoidal signal is shown by the
solid line 
in the middle panel.  From the rate curve, we 
derive a corresponding counting series with a time step 2 ms and
use 8192-point FFTs to get the Fourier power densities $p_f(f_j)$ and the 
corresponding 
densities in the time area $p_f(\Delta t_j=1/f_j)=p_f(f_j)f_j^2$.
The Fourier spectrum is shown by the dashed line in the middle panel of Fig. 1. 
The areas under the two power density curves in Fig. 1 
are the same, as dictated  by Parseval's theorem. It is obvious that the
Fourier spectrum
can not be interpreted as the distribution of variability amplitude vs. time
scale.
The power being concentrated at the sinusoid frequency $f$ does not mean that 
the intensity variation of the process exists only at the corresponding time
scale $1/f$.
From the light curves with time steps $\Delta t=0.05$ s and 0.28 s, shown in
the bottom panel of Fig. 1, we can see that variation of the intensity
definitely
exists at time scales $\Delta t < 0.5$ s, but almost no power does in the
Fourier
spectrum at those time scales. In contrast with the Fourier spectrum, 
the power density spectrum 
determined in the time domain (the solid line in the middle panel of Fig.1)
gives a proper description of the variability amplitude distribution of time
scales.

Figure 2 shows the power density spectrum (the solid line in the bottom panel) 
derived in the time domain and the corresponding Fourier spectrum 
(the dashed line in the bottom panel) 
for a periodic triangular signal (the top panel).
The Fourier spectrum is obtained with 8192-point FFT for a light curve with
time step 1 ms. The many peaks at short time scales in the Fourier
spectrum (high frequency harmonics) do not mean that there really exist
strong variations in the process at those time scales, which  
are just necessitated by mathematically decomposing the triangular signal into
sine waves.
As a result, a Fourier spectrum may overestimate power densities 
at short time scales (high frequency harmonics region) for 
a periodic signal with pulse shape being far from a sinusoid.	 

(2) {\sl Stochastic Shots}

We now consider the power spectra of a signal $s(t)$ consisting of stochastic
shots. 
Both the rise and decay fronts of the shots are an exponential with a time
constant 
$\tau$ taken uniformly from the range between 5 ms and 0.2 s.
 The separation 
between two successive shots is exponentially
distributed with an average separation $d=0.5$ s. The average signal rate is 
assumed as $\bar{r}_s = 300$ cts/s. The peak height of the shot 
follows a uniform distribution between zero and the maximum.
The top panel of Fig. 3 shows a piece of the signal counting curve with time
step
$\Delta t=0.01$ s and the solid line in the bottom panel of Fig. 3 is the  
power density spectrum in the time domain expected for the signal $s(t)$. 
A simulated 2000 s light curve $x(k)$ with 1 ms time step 
is produced by a random sampling of the signal curve with Poisson fluctuations
plus
a white noise with mean rate 5000 cts/s. The middle panel of Fig. 3 shows a
piece
of the light curve with $\Delta t=0.01$ s obtained from the simulated 1 ms 
light curve. In calculating the signal power density at a time scale $\Delta
t$, 
The total light curve with time step $\Delta t$ was divided into $M$ data
segments  
with $N=100$ bins each. If the segments number $M<100$ in cases of 
large time scales, we let $M=100$ and decreased the number $N$ of data points
in each 
segment accordingly. 
For each segment of the light curve with time bin $\Delta t_1$, the total
powers
at two time scales $\Delta t_1$ and $\Delta t_2=2\Delta t_1$ through Eq.~(5)
were
calculated.  The corresponding noise powers were
calculated by Eq.~(8) with $r$ set equal to the average counting rate.
The total power density $p(\Delta t)$ and noise power density 
$p_{noise}(\Delta t)$ at $\Delta t=1.5\Delta t_1$  can be calculated 
by Eq.~(7) and (9), respectively. Finally, the signal power density in the time 
domain is 
$p_{signal}(\Delta t)=p(\Delta t)-p_{noise}(\Delta t)$.
In the bottom panel of Fig.~3, the plus signs mark the average signal power
densities 
at different time scales.
For the same light curve with 1 ms time binning, we also calculated the Leahy
density 
$w(f_j)=2|a_j|^2/a_0$ for each $T=4.096$~s segment, where $a_j$ is the Fourier
amplitude at frequency 
$f_j=j/T$ determined from a 4096-point FFT. It is well known that the noise
Leahy density 
$w_{noise}=2$,
 so the signal Leahy density can be written as $w_{signal}(f_j)=w(f_j)-2$ and 
the Fourier power density of the signal is expressed by  
$p_{F,signal}(f_j)=w_{signal}(f_j)a_0/T$ (\cite{lea83}, \cite{van88}). 
The dashed line in the bottom panel of Fig. 3 
shows the average Fourier power density spectrum of the signal with respect to 
time scale $\Delta t_j=1/f_j$. 
As the characteristic time $\tau$ of a shot is taken in the range between 5 ms
and 0.2 s, 
there should exist considerable variability over this time scale
range as represented by the power density spectrum in the time domain expected
for
the signal $s(t)$ (the solid line in the bottom panel of Fig.~3) and by 
that obtained 
from the simulated data including noise (pluses in the bottom panel of Fig.~3). 
Figure~4 represents the power spectra for another shot model with shorter
characteristic time
constant $\tau$ between 0.5 ms and 2 ms, average separation
between two successive shots 3 ms, average signal rate 
$\bar{r}_s=500$ cts/s, and noise rate 5000 cts/s. 
>From Fig.~3 and Fig.~4 we can see that for a random shot series
the Fourier spectrum is more or less consistent with the power spectrum
in the time domain at time scales greater than the characteristic time scale
 of the model, 
but significantly underestimates the power densities at shorter time scales.

(3) {\sl Markov process}

The Markov process or autoregressive process  of first order can describe 
the character of the variability for many physical processes. A Markov process 
can be expressed as the following 
stochastic time series	  
\[ u(k)=a\cdot u(k-1)+\epsilon ~,\]
where $\epsilon$ is a Gaussian random  variable with zero mean and unit
variance,
and the relaxation time of the process is $\tau = -\Delta t/\log | a |$
with $\Delta t$ being the time step.
The observed light curve for the signal is 
\[ s(k)=c\cdot u(k) + r_s\Delta t~, \]
where $r_s$ is the average rate of the signal.
We make a light curve of the signal $s(k)$ with $\Delta t=0.01$~s, $\tau =
0.1$~s,
$r_s=2000$~cts/s and $c=2.5$. A piece of the produced signal light curve is
shown
in the top panel of Fig.~5.  
The final observed light curve $x(k)=s(k)+n(k)$ with $n(k)$ being 
a Poisson noise 
with mean rate 5000~cts/s, is shown in the middle panel of Fig.~5. 
In the bottom panel of Fig.~5, the solid line
shows the power density distribution expected for the signal $s(k)$,
the plus signs indicate excess power densities in the light curve $x(k)$ 
estimated by Eqs.~(5)-(10) and the dashed line by FFT. 
From Fig.~5, we can see, similar to the case of stochastic shot process, that
the 
proposed algorithm of evaluating power densities in the time domain is capable
of extracting the
power spectrum of the signal from noisy data and that  
the Fourier spectrum significantly underestimate
the power densities at time scales around or shorter than the characteristic
time scale of a stochastic process.  
 
\section{ACCRETING NEUTRON STARS AND BLACK HOLES}

As revealed by above study based on simulations, the power density spectrum
derived in the
time domain can  describe the real power distribution with respect to
time scale for different time processes. The two kinds of power spectrum, 
Fourier spectrum and spectrum in the time domain, differ in ways
depending on models of the processes.
By comparing the two power spectra, more information about the nature
of the variability of an object can be extracted.
For this purpose we calculate both the power density spectra in the time domain
and 
Fourier spectra for the publicly available data of the Proportional 
Counter Array (PCA) 
aboard the Rossi X-ray Timing Explorer (RXTE) for a sample of X-ray binaries,
7 neutron stars and 7 black hole binaries.
The PCA observations of X-ray objects included in our study are listed in Table
1. 
 For each analyzed observation, we use the version 4.1 of standard RXTE ftools
to
extract the PCA data.  
At 18 time steps between 0.001 s and 2.5 s, we make the corresponding light
curves
with $\sim 2000$ s duration in an energy band as noted in Table 1. Then we
remove all  
ineffective data points caused by failure due to the satellite, detector or
data 
accumulation system, and calculate 
the power density spectra in the time domain, being based on a similar
procedure in our simulation.  
The corresponding Fourier spectra are constructed by using $2^{-12}$ s ($\sim
244~\mu s$)
time resolution light curves divided into parts containing 8192 bins.

Figure 6 shows the results from 6 observations for 5 neutron star binaries: 
KS 1731-260, 4U 1705-44, GS 1826-24, 4U 0614+091, 4U 1608-522 in the low state
and
4U 1608-522 in the high state. The two kinds of power spectrum in the studied
accreting neutron stars are generally consistent with each other, at least for
the continuum dominated region.  The feature 
is rather complicated for such sources whose Fourier spectra
have significant quasi-periodic oscillation (QPO) structure. The left panel 
of Fig. 7 is from the QPO source, Sco X-1, whose power density 
spectrum in the time domain also shows a strong QPO structure but the  
peak shifts to shorter time scales due to the steep slope of the
sinc function near the Nyquist frequencies (7 - 12.5 Hz) as more and
more of the 6+ Hz QPO is accommodated by the time sampling. 
The difference between the spectrum of time domain and the 
Fourier spectrum at short time scales for Cyg X-2 (see the 
right panel of Fig. 7) may also be caused by QPOs.

For black hole candidates, we first analyze the canonical source Cygnus X-1. 
On May 10, 1996 (day 131 of 1996), the All-Sky Monitor on RXTE revealed 
that Cyg X-1 started a transition from the normal low (hard) state 
to a high (soft) state. After reaching the high state, it stayed there 
for about 2 months before going back down to the low state (\cite{cui97}).
During this period, 11 pointing observations of Cyg X-1 were made by RXTE. 
We use one observation of PCA/RXTE for each of the four states of Cyg X-1. 
The signal power density spectra in the time domain of Cyg X-1 are shown by the
plus
signs in Fig. 8 and the corresponding Fourier spectra by the dots in Fig. 8.
The power spectra of Cyg X-1 in different states (Fig. 8) have a
common trait that  the Fourier spectra are significantly lower than the
corresponding
power spectra of time domain in the time scale region of $\Delta t<0.1$~s.

Besides Cyg X-1, the black hole candidates 
GRS1915+105, GRO J1655-40, GX 339-4, and XTE J1550-564 also demonstrate a
significant
excess in the power spectra in the time domain in comparison with the
corresponding Fourier spectra at time scales shorter than $\sim 0.1$ s
(see Figure 9). But the another analyzed black hole binary, GRS 1758-258, 
with QPO structures in its Fourier spectrum behaves differently 
(Fig. 10). 

\section{DISCUSSION} 
 
A Fourier power density spectrum presented in the time domain can not be  
interpreted as the real power density distribution of the physical process 
studied. 
In principal, the power densities in the time domain can be derived from a
Fourier 
spectrum only when one knows every power density spectrum in the time domain
for each sinusoidal function at all Fourier frequencies and adds them up with
weight factors being the Fourier amplitudes. 
We propose here studying power spectra directly in the time domain.	
The definition of the power $P(\Delta t)$ (Eq. 5) 
is based  just on the 
original meaning of rms variation and the power density spectrum $p(\Delta t)$ 
(Eq. 6) represents the distribution of the variability amplitude  
vs. time scale. 
The power density spectrum $p(\Delta t)$ in the time domain obtained from 
an observation depends only on the intrinsic nature of the signal process   
and the statistical property of the observed data, as does the Fourier 
power spectrum in the frequency domain. Our simulation studies show
that
the proposed algorithm, Eqs. (7), (9) and (10), is capable of extracting 
power densities of the signal from noisy data (comparing the expected signal
spectra,
the solid lines in the bottom panels of Figs. 3 and 5 and in Fig. 4, with 
the spectra calculated from the noisy data, plus signs in corresponding plots).
These results indicate that the technique of spectral analysis in the time
domain
is a useful tool in timing and worth applying in temporal analysis
for different sources.	 
From Figs. (1) - (5) one can see that the difference of
the Fourier spectrum with the spectrum in the time domain is dependent on 
the model of time series and sensitive to the characteristic time
of a stochastic process. We can then use the difference between two kinds of
spectrum to study the intrinsic nature
of a studied process.

The power spectra shown in Figs. (8) and (9) for black hole candidates 
have a common character that 
the Fourier spectra are significantly lower than the corresponding 
time-based spectra in the time scale region of $\Delta t<0.1$~s, which	is
similar 
to the simulation results for the stochastic shot model (Fig. 3) or 
the Markov process (Fig. 5). The existence of stochastic shots in X-ray light
curves
of Cyg X-1 has been noticed for a long time. With an improved searching and
superposing 
algorithm and PCA/RXTE data of Cyg X-1 in different states, Feng, Li \& Chen
(1998) 
find that the average shot profiles can be described by exponentials with 
characteristic time scales $\sim 0.01 - 0.1$ s. Pottschmidt et al (1998)    
point out that an autoregressive process of first order with a relaxation time 
$\tau$ of about 0.1~s can reproduce approximately the
variability of Cyg X-1. Thus, the characteristics of time process
in Cyg X-1 revealed from our power spectral analysis is consistent with 
that from modeling its light curves. The black hole candidates 
GRS 1915+105, GRO J1655-40, GX 339-4 and XTE J1550-564 also demonstrate 
characteristics similar to Cyg X-1, indicating that a stochastic process
with
a characteristic time $\sim 0.1$ s may be common in accreting black holes.
While the absence of the broad-band noise above approximately 100 Hz in black
hole 
candidates has been noticed before (\cite{sun00}), our analysis shows that
the apparent absence in Fourier spectra is caused by the existence of 
a stochastic process with characteristic time $\sim 0.1$ s and by
the insensitivity of the Fourier technique to
detecting rapid variability in a stochastic process.
          
At the same time, the two kinds of power spectrum are more or less consistent
for accreting neutron stars with continuum dominated Fourier spectrum, 
as shown in Fig. 6. 
Our simulation results, Figs. (3) - (5), show that for a stochastic process 
the two kinds of spectrum can be consistent with each other at 
time scales greater than the characteristic time of the process.  
Assuming that a significant variability of an accreting neutron star comes 
from a stochastic process with very short characteristic time constant 
$\tau\ll 1$ ms can explain the consistence of the two kinds of 
power spectrum observed in neutron star systems.   
Sunyaev \& Revnivtsev (2000) find that the power density spectra of accreting
neutron stars with a weak magnetic field have significant broad noise component
at the frequency 500-1000 Hz.  They suggest that 
those X-ray transients which demonstrate significant noise in their X-ray flux
at frequencies above $\sim 500$ Hz should be considered neutron stars.
Most sources studied in this work (neutron star systems 4U 0614+091, 4U
1608-522, 
GS 1826-24, 4U 1705-44, KS 1731-260, Cyg X-2, and black hole candidates Cyg
X-1,
GX 339-4, GRS 1915+105, GRO J1655-40, XTE J1748-288, GRS 1758-258) are also
 studied by Sunyaev \& Revnivtsev (2000) and our results 
support their claim under the condition that
the spectra and noise in their statement are restricted within the Fourier
framework.
In contrast to the results from Sunyaev \& Revnivtsev (2000), 
our results are based on inferring the characteristic time
from the relation between two kinds of spectrum, no matter what the absolute 
magnitude of power spectrum is. The characteristic feature we find appears in
all spectral states of
the black hole candidate Cyg X-1 and the same is true with the neutron star 
binary 4U 1608-522.  

The characteristic features in rapid variability for two different 
kinds of X-ray binary in our sample, 
which associate significant stochastic processes at the time scale 
$\sim 0.1$ s for black hole 
candidates and $\ll 1$ ms for neutron stars, are revealed only in
continuum or broad noise. The QPO components behave in a more complicated
fashion, but can be understood through the sensitivity of the
time-based spectrum to variation at 2$\delta$t, i.e., at half
the Fourier sampling frequency (the Nyquist frequency). 
It has been found that QPO structures in power spectra can be caused by
different kinds of signal, e.g., modulated periodic
signals and stochastic autoregressive processes with order $\ge 2$
(\cite{van86}).  
In the case of the neutron star binary Sco X-1 (the left
plot of Fig. 7), the power density spectrum in the 
time domain reveals the QPO feature surprisingly well, though with a peak
shifting to
shorter time scales and with worse resolution. 
  
Different processes with essentially  different natures could result in almost
the 
same Fourier power spectrum, thus  distinguishing them is difficult through 
timing analysis only with the Fourier technique.  This is a reason why 
we need to develop and apply alternative methods to supplement the Fourier
technique
in spectral analysis. Our results, 
though only preliminary, show that simultaneous use of both the Fourier and 
the time domain methods can help in probing the intrinsic nature
of timing phenomena, and further, in distinguishing
between different kinds of accreting compact object.	 
     
The authors thank the referee for helpful comments and suggestions and Dr. Qu
Jinlu 
for help in data treatment. This work is supported
 by the Special Funds for Major State Basic Research Projects and
the National Natural Science Foundation of China. The data analyzed in this
work are obtained through 
the HEASARC online service provided by the NASA/GSFC.

\clearpage
\begin{deluxetable}{crrrr}
\footnotesize
\tablecaption{The used PCA observations of X-ray binaries \label{tbl-1}}
\tablewidth{0pt}
\tablehead{
\colhead{Type} & \colhead{Object} & \colhead{Observation}& \colhead{Band
(keV)}& 
\colhead{Note}
} 
\startdata
 &KS 1731-260&10416-01-02-00 &3-21 &~ \nl
 &4U 1705-44&20073-04-01-00 &3-20 &~ \nl
 &GS 1826-24&30054-04-01-00 &3-20 & \nl
Neutron Star &4U0614+091&30054-01-01-01 &3-21 & \nl
 &4U1608-522&30062-01-01-04 &3-21 &low state\nl
 &4U1608-522	 &30062-02-01-00&3-21 &high state\nl
 &Sco X-1&30035-01-02-000 &2-18 & \nl
 &Cyg X-2&30418-01-01-00 &2-21 & \nl
\cline{1-5}
& Cyg X-1  &10412-01-01-00 & 2-13&low to high\nl
  &Cyg X-1	  &10512-01-08-00 &2-13&high state\nl
  &Cyg X-1	  &10412-01-05-00 &2-13&high to low\nl
  &Cyg X-1	  &10236-01-01-03 &2-13&low state\nl
Black Hole&GRS1915+105&20402-01-05-00 &5-22 & \nl
&GRO J1655-40 &20402-02-25-00 &5-22& \nl
& GX 339-4 &20181-01-01-00 &4-22 & \nl
&XTE J1550-564 &30191-01-14-00 &2-13 & \nl
&GRS 1758-258&30149-01-01-00 &3-21 & \nl
&XTE J1748-288&30185-01-01-00 &2-21& \nl 
\enddata
\end{deluxetable}

\clearpage
\begin{figure}
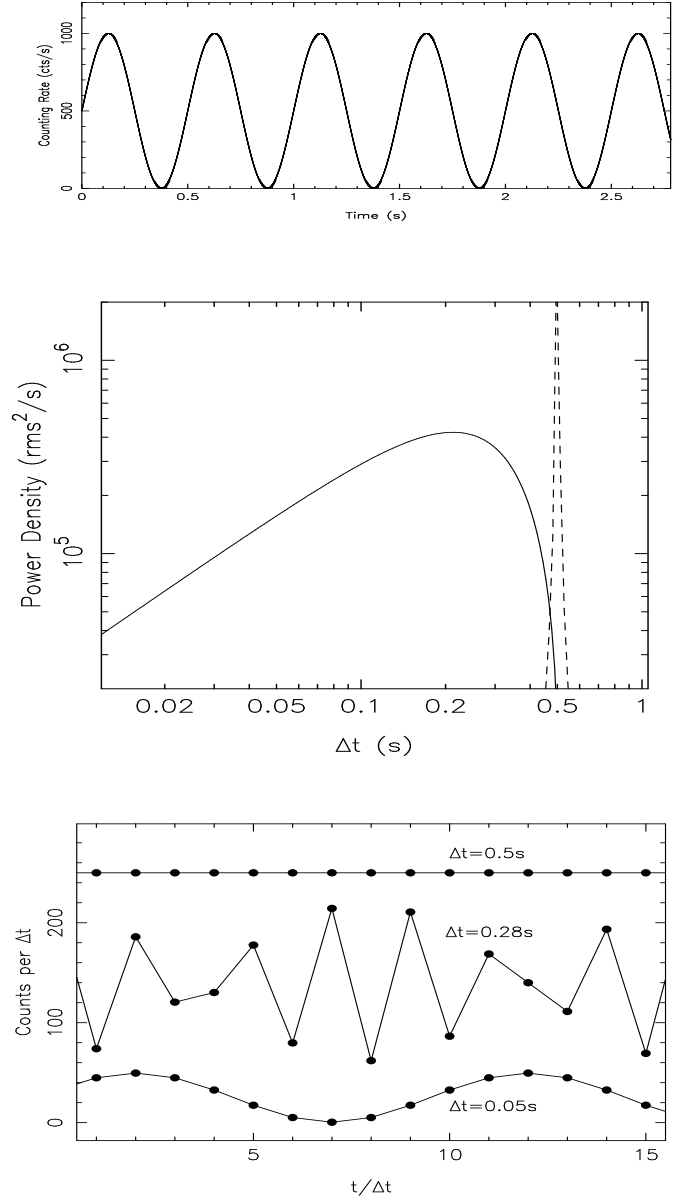

\epsscale{1.0}
\vspace{5cm}
\plotfiddle{f1a.eps}{20pt}{-90}{40}{25}{-200}{200}
\vspace{4cm}
\plotfiddle{f1b.eps}{20pt}{-90}{50}{35}{-190}{200}
\vspace{5.5cm}
\plotfiddle{f1c.eps}{20pt}{-90}{40}{30}{-200}{200}
\vspace{-1cm}
\caption{Distribution of power density vs. time scale of a sine signal.
 {\sl Top panel}: counting rate curve of the sine signal with period 0.5 s. 
 {\sl Middle panel}: power densities. {\it Solid line} -- power density
spectrum derived in the time domain. {\it dashed line} -- Fourier spectrum, 
derived by 8192-point FFT for the sine signal light curve with step 2 ms
 and shifted downwards	by a factor of 100. {\sl Bottom panel}: three light
curves with
time step 0.05 s, 0.28 s, and 0.5 s, respectively.
\label{fig1}}
\end{figure}

\clearpage

\begin{figure}
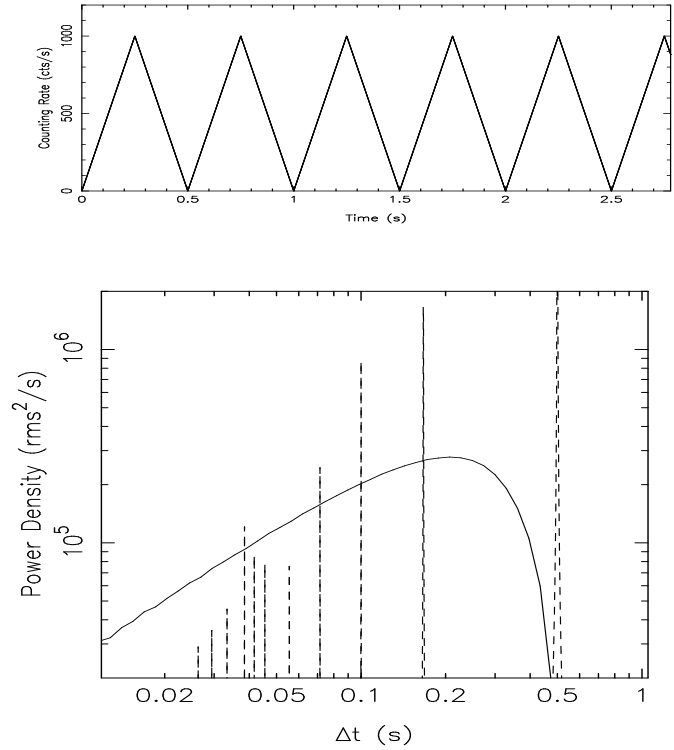

\epsscale{1.0}
\vspace{4cm}
\plotfiddle{f2a.eps}{12pt}{-90}{40}{25}{-200}{200}
\vspace{4cm}
\plotfiddle{f2b.eps}{12pt}{-90}{50}{35}{-190}{200}
\caption{Distribution of power density vs. time scale of a periodic
triangle signal.
 {\sl Top panel}: counting rate curve of the periodic triangular signal
with period 0.5 s. 
 {\sl Bottom panel}: power densities. {\it Solid line} -- power density
spectrum derived in the time domain. {\it dashed line} -- Fourier spectrum, 
derived by 8192-point FFT for the the light curve of the signal with 1 ms step
and
shifted downwards by a factor of 80. 
\label{fig2}}
\end{figure}

\clearpage

\begin{figure}
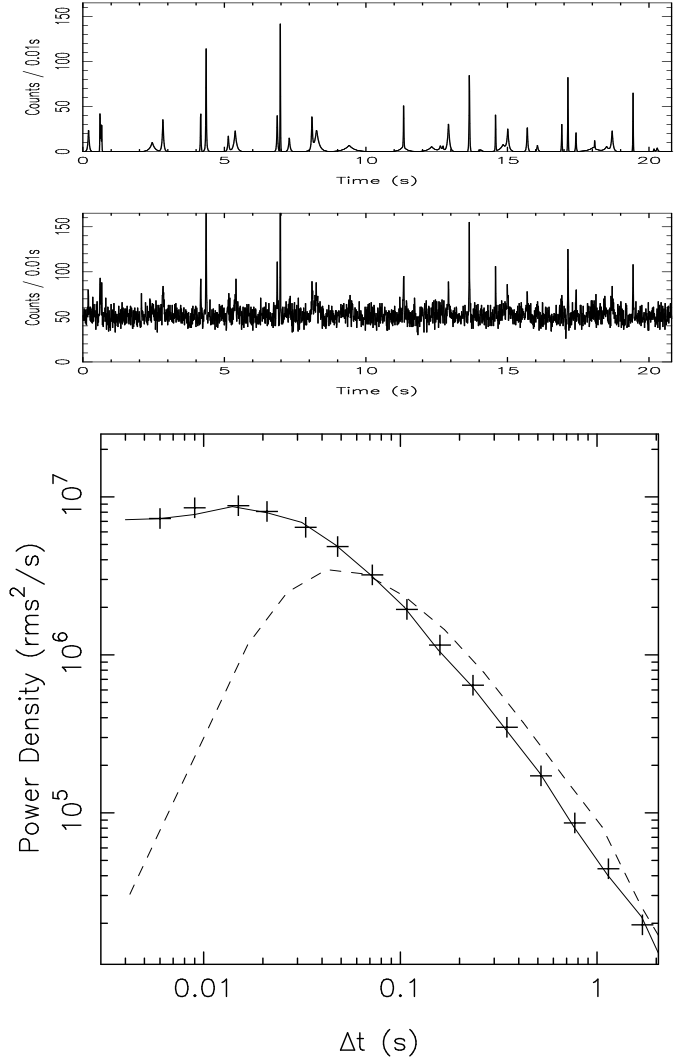

\epsscale{1.0}
\vspace{5cm}
\plotfiddle{f3a.eps}{20pt}{-90}{40}{20}{-200}{220}
\vspace{1.5cm}
\plotfiddle{f3b.eps}{20pt}{-90}{40}{20}{-200}{220}
\vspace{-0.9cm}
\plotfiddle{f3c.eps}{20pt}{-90}{51}{48}{-190}{130}
\vspace{5cm}
\caption{Distribution of power density vs. time scale of a shot model. 
{\sl Top panel}: the signal, stochastic exponential shots with time constant
$\tau$ 
between 5 ms and 0.2 s. {\sl Middle panel}: simulated data which involved both
 signal and Poisson noises.
{\sl Bottom panel}: signal power densities . {\it Solid line} -- power density
 distribution of time scale expected for the signal. 
{\it Dashed line} -- excess Fourier spectrum from the simulated data.
{\it Plus signs} -- excess power densities 
calculated by the timing technique in the time domain for the simulated data. 
\label{fig3}}
\end{figure}

\clearpage

\begin{figure}
\epsscale{1.0}
\vspace{5cm}
\plotfiddle{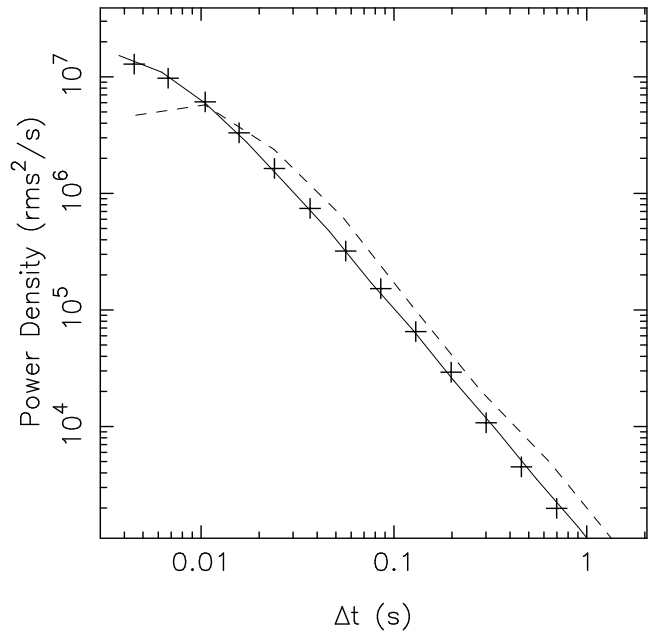}{20pt}{-90}{50}{48}{-190}{130}
\vspace{5cm}
\caption{Distribution of power density vs. time scale of a shot model. Time
constants of exponential shots are between 0.5 ms and 2 ms.  
{\it Solid line} -- a theoretical distribution
of power densities expected for the signal. 
{\it Dashed line} -- the excess Fourier spectrum from the simulated data.
{\it Plus} -- the excess power densities
calculated in the time domain for the simulated data. 
\label{fig4}}
\end{figure}

\clearpage

\begin{figure}
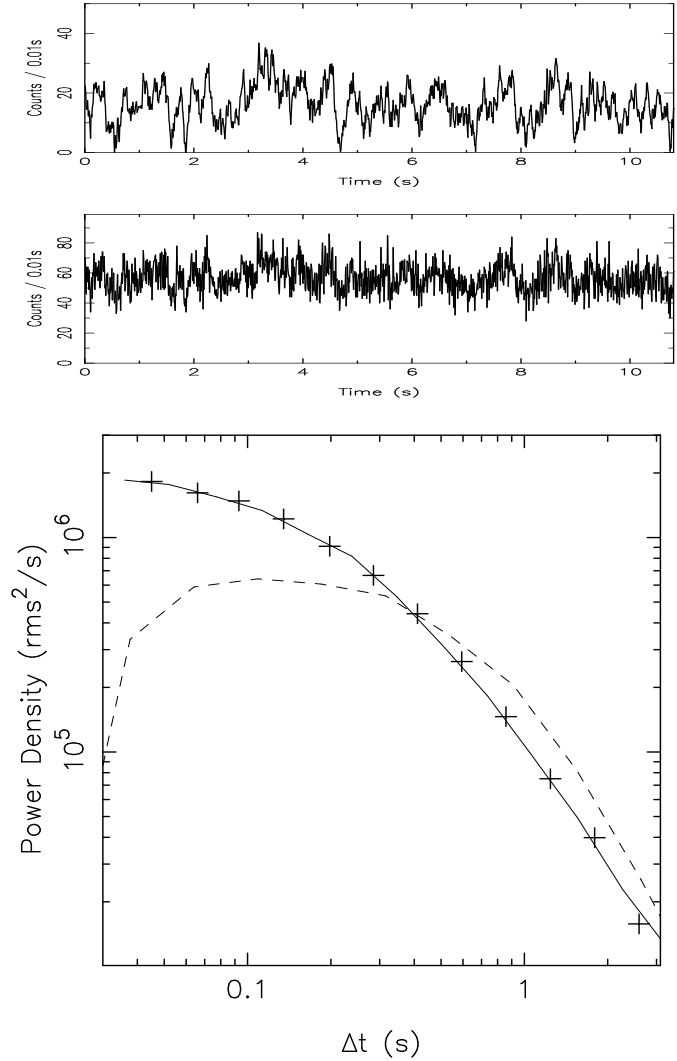

\epsscale{1.0}
\vspace{5cm}
\plotfiddle{f5a.eps}{20pt}{-90}{40}{20}{-200}{220}
\vspace{1.5cm}
\plotfiddle{f5b.eps}{20pt}{-90}{40}{20}{-200}{220}
\vspace{-0.9cm}
\plotfiddle{f5c.eps}{20pt}{-90}{51}{48}{-190}{130}
\vspace{5cm}
\caption{Distribution of power density vs. time scale of a Markov process. 
{\sl Top panel}: the signal, a Markov process with relaxation time 0.1 s.
 {\sl Middle panel}: simulated data which include  signal and Poisson noises.
{\sl Bottom panel}: power densities of the signal. {\it Solid line} -- the 
power distribution of time scale expected for the signal. 
{\it Dashed line} -- excess Fourier spectrum from the simulated data.
{\it Plus} -- excess power densities
calculated in the time domain for the simulated data. 
\label{fig5}}
\end{figure}

\clearpage

\begin{figure}
\vspace{4.55cm}
\epsscale{1.0}
\plotfiddle{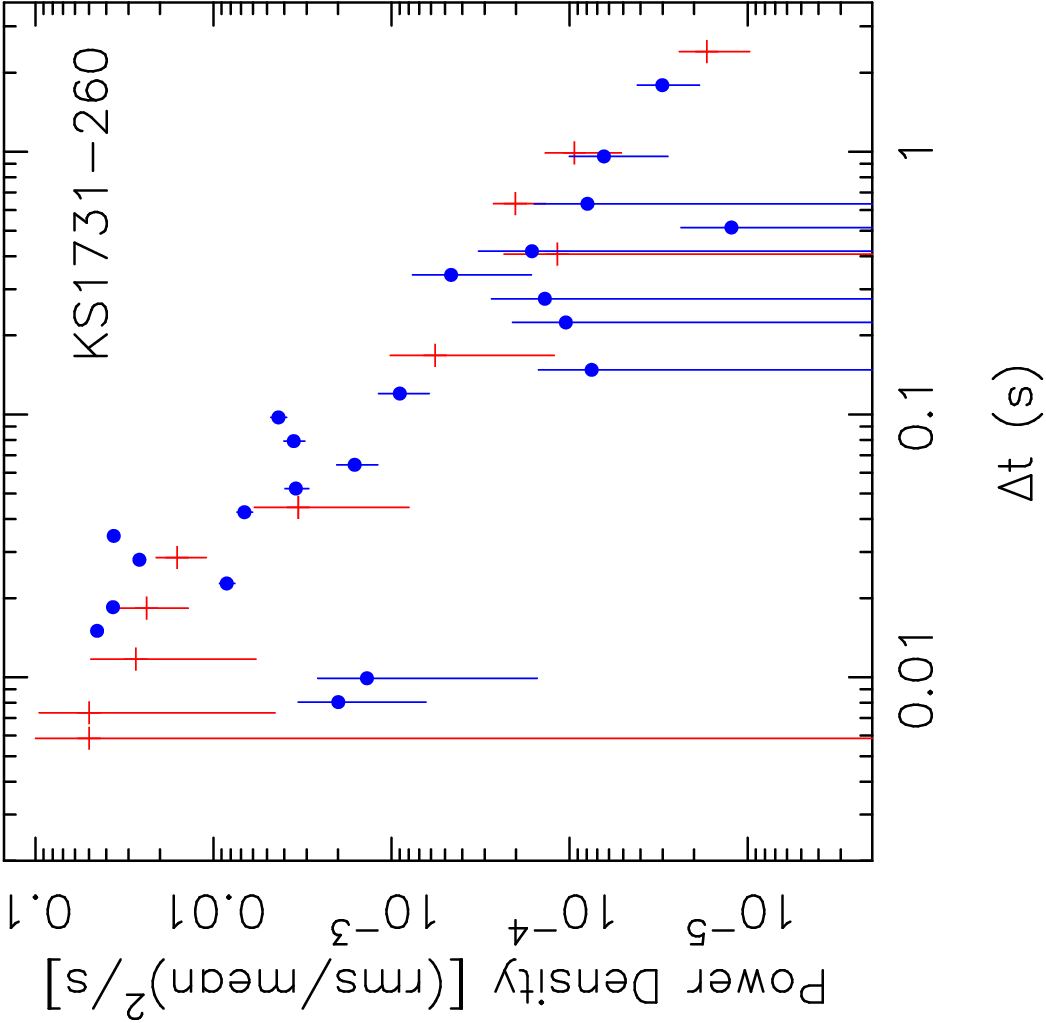}{20pt}{-90}{40}{25}{-260}{178}
\plotfiddle{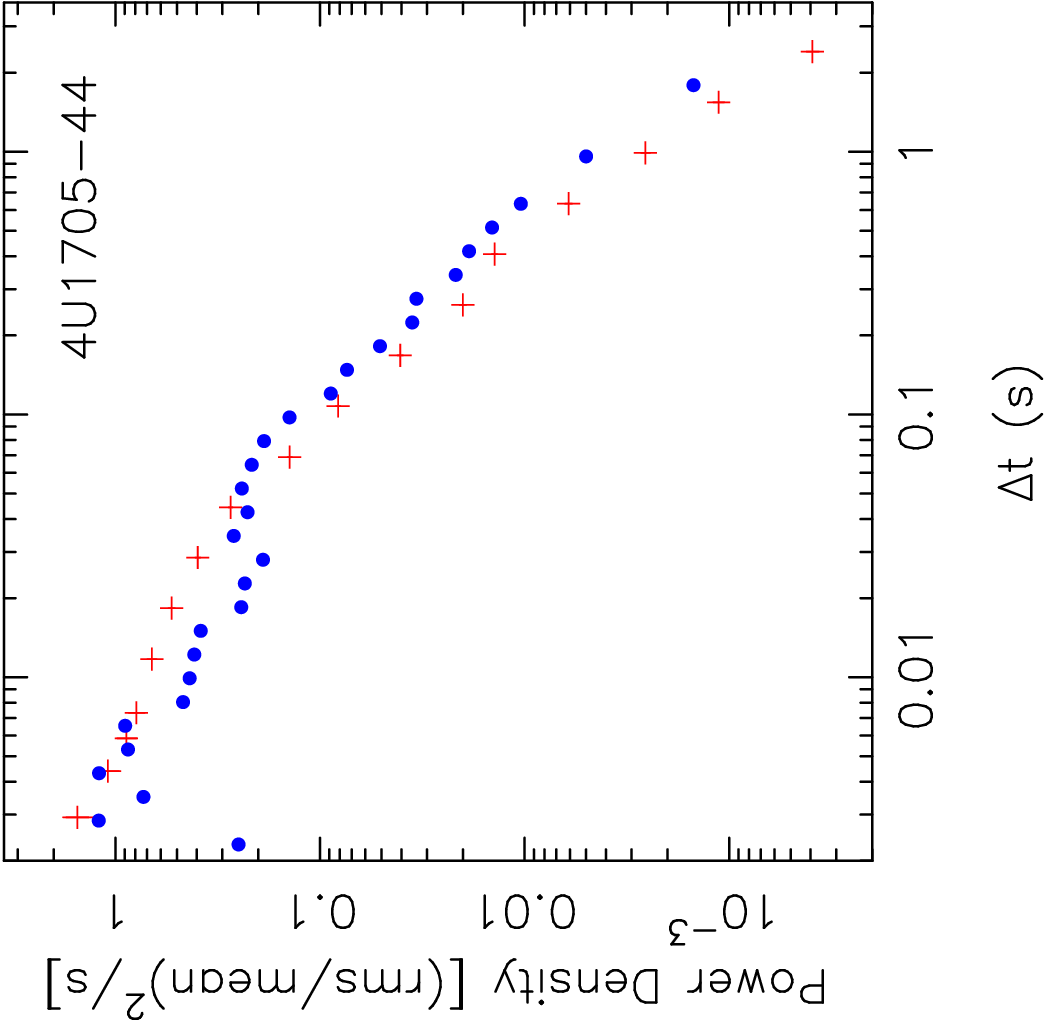}{20pt}{-90}{40}{25}{-50}{215}
\plotfiddle{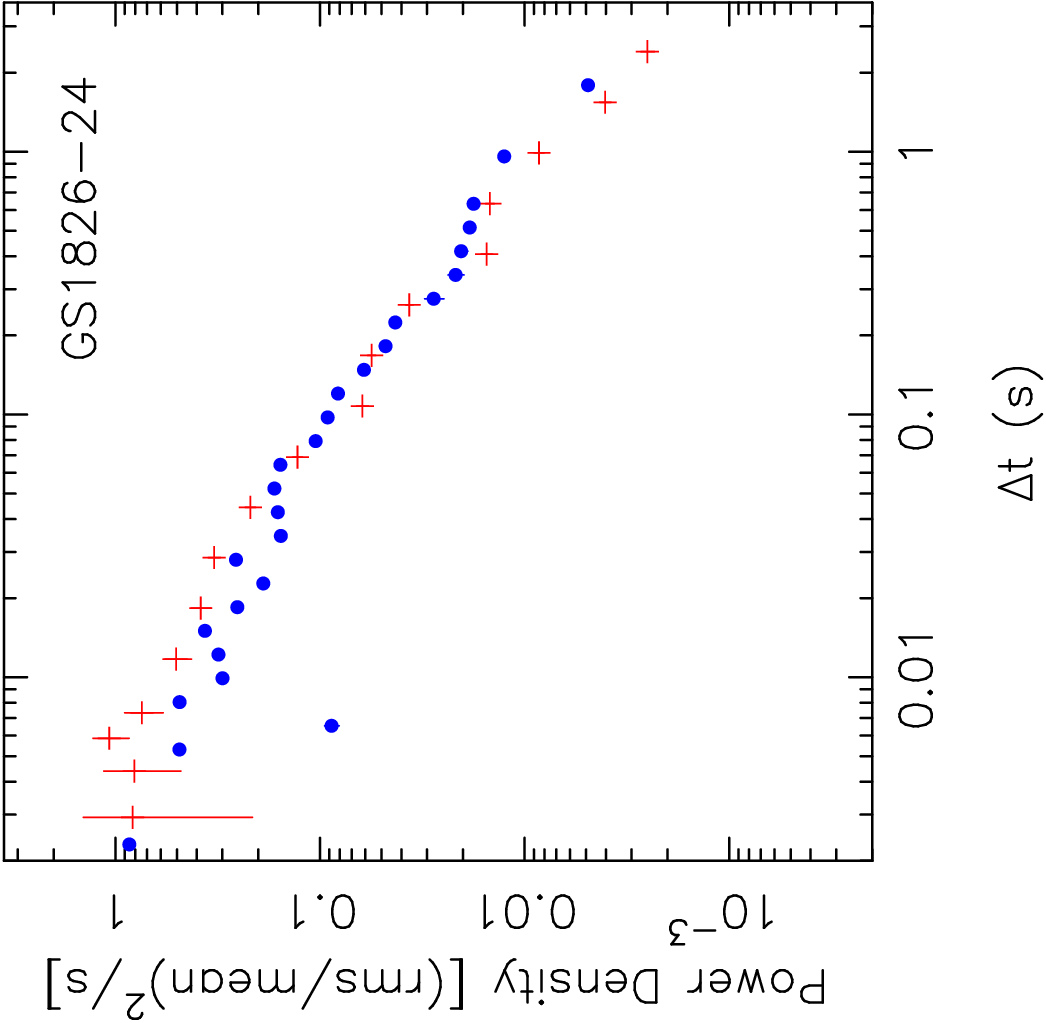}{20pt}{-90}{40}{25}{-260}{119}
\plotfiddle{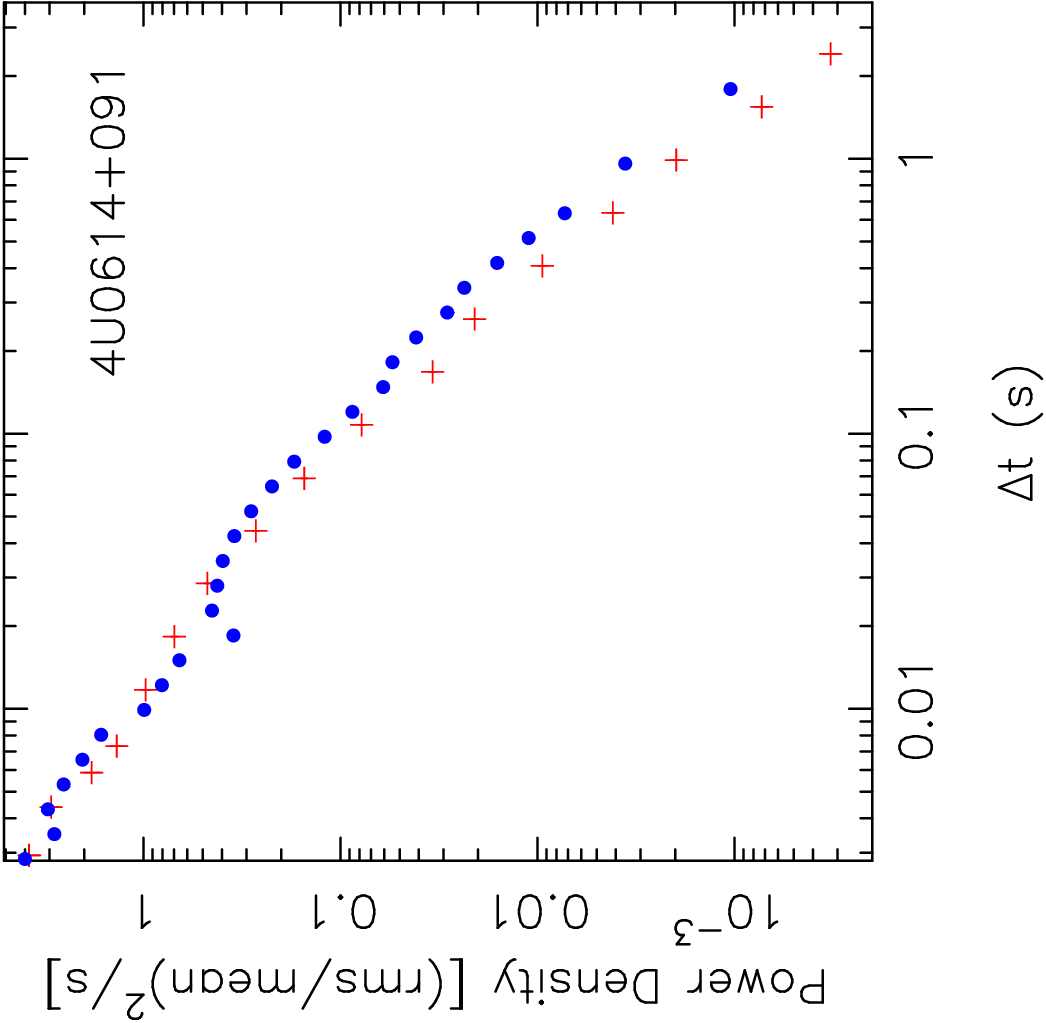}{20pt}{-90}{40}{25}{-50}{155}
\vspace{20mm}
\plotfiddle{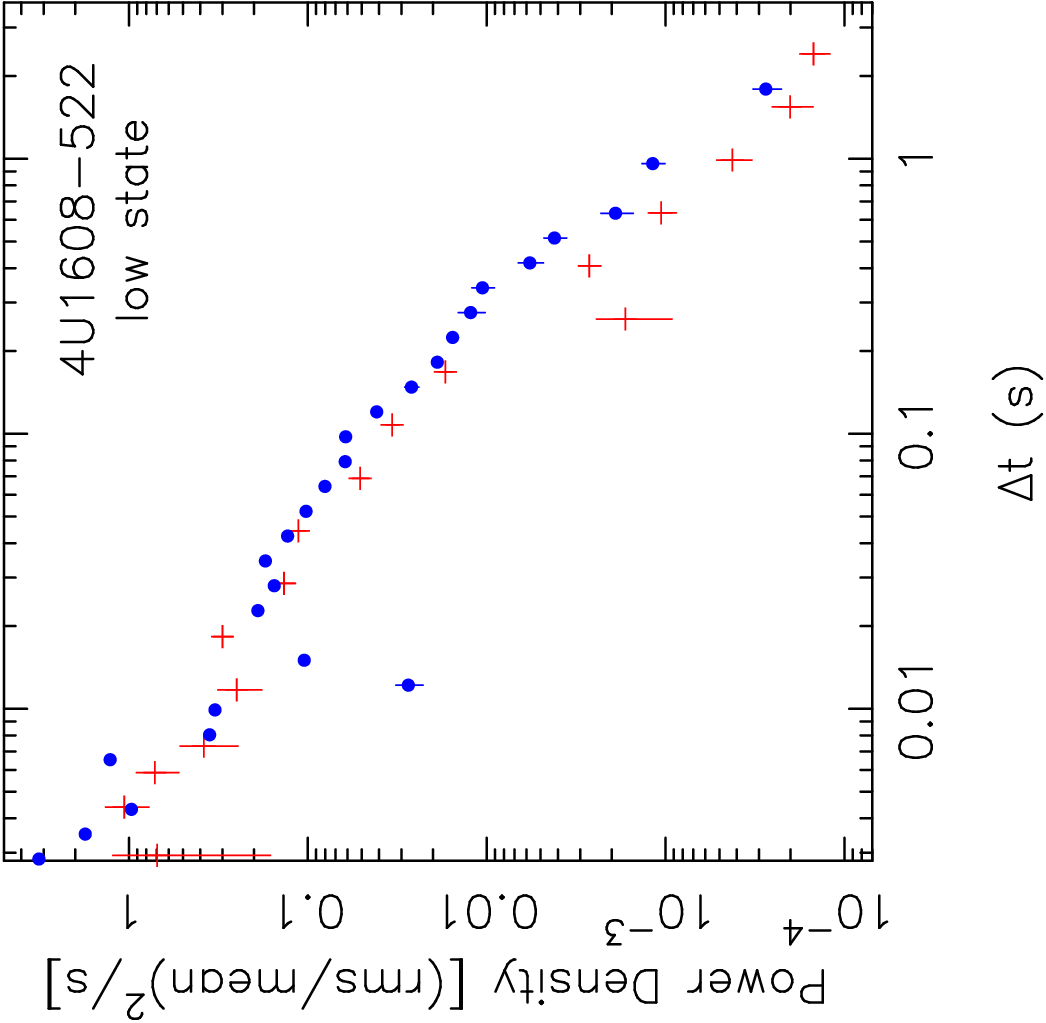}{20pt}{-90}{40}{25}{-260}{119}
\plotfiddle{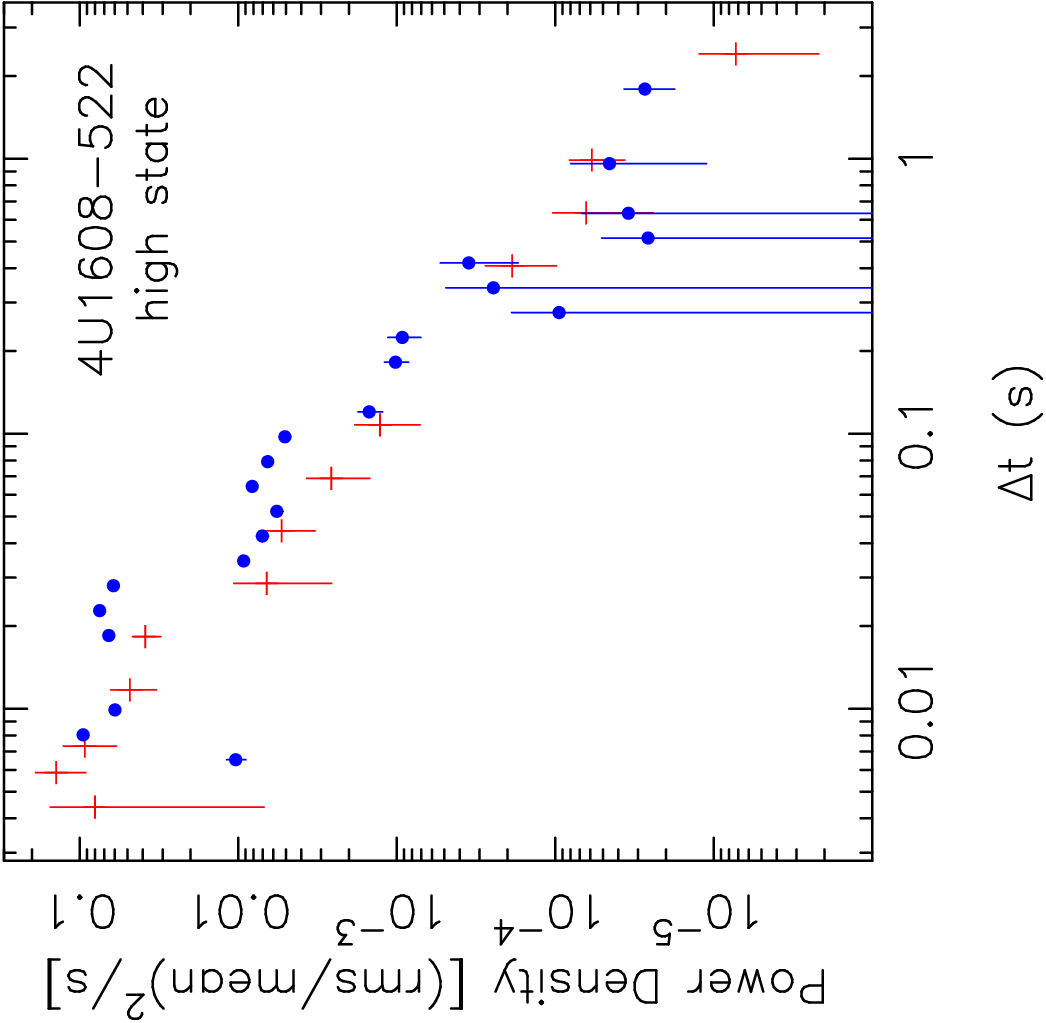}{20pt}{-90}{40}{25}{-50}{155}
\vspace{-4mm}
\caption{Power density vs. time scale of 5 accreting neutron stars.
{\it Plus}: spectrum in the time domain. {\it Dot}: Fourier
spectrum. 
\label{fig6}}
\end{figure}

\clearpage

\begin{figure}
\vspace{5cm}
\epsscale{1.0}
\plotfiddle{f7a.eps}{20pt}{-90}{40}{25}{-260}{178}
\plotfiddle{f7b.eps}{20pt}{-90}{40}{25}{-50}{215}
\vspace{-3cm}
\caption{Power density vs. time scale of two accreting neutron stars.
{\it Plus}: spectrum in the time domain. {\it Dot}: Fourier
spectrum. 
\label{fig7}}
\end{figure}


\begin{figure}
\vspace{2cm}
\epsscale{1.0}
\plotfiddle{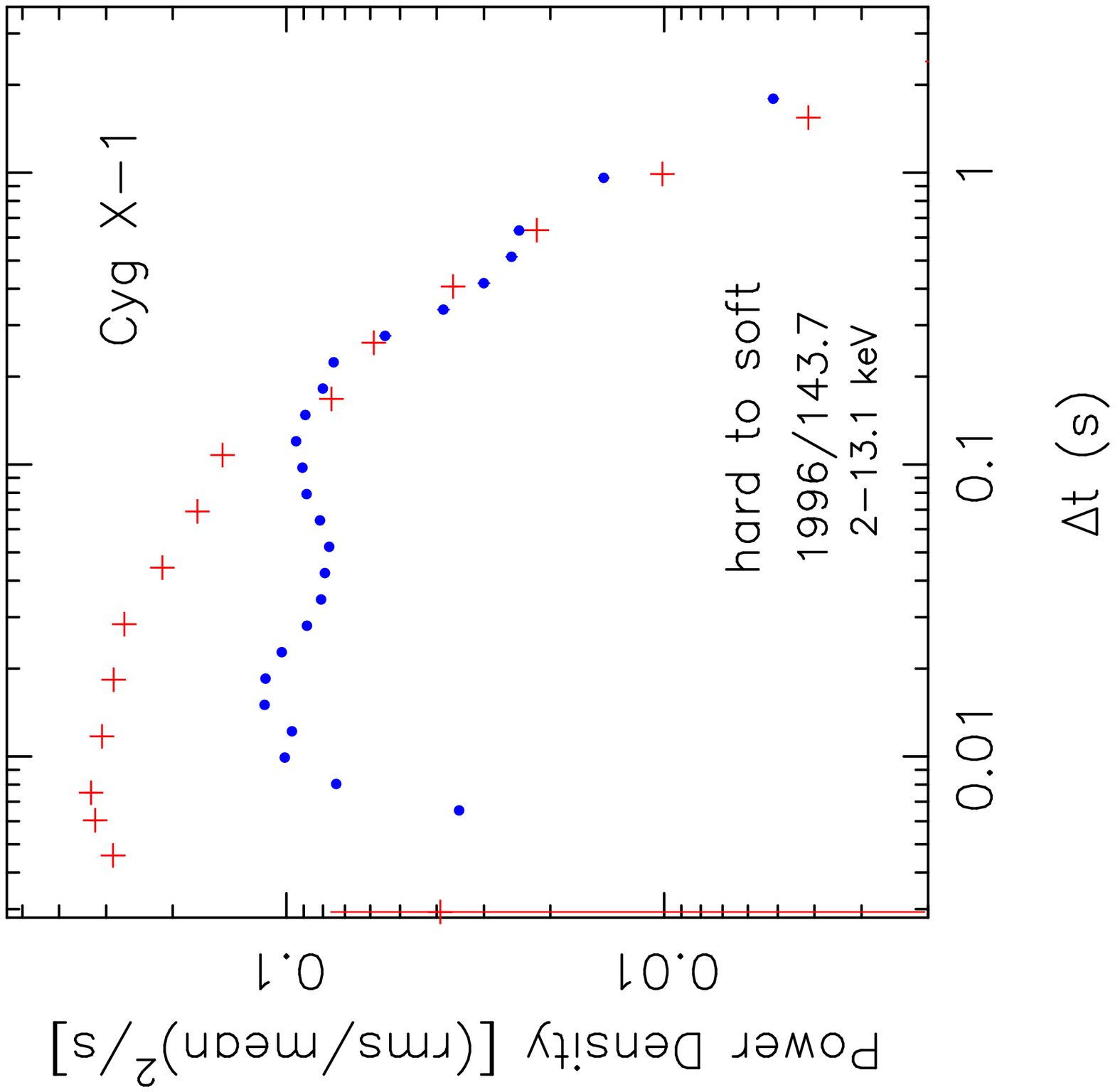}{20pt}{-90}{40}{25}{-260}{178}
\plotfiddle{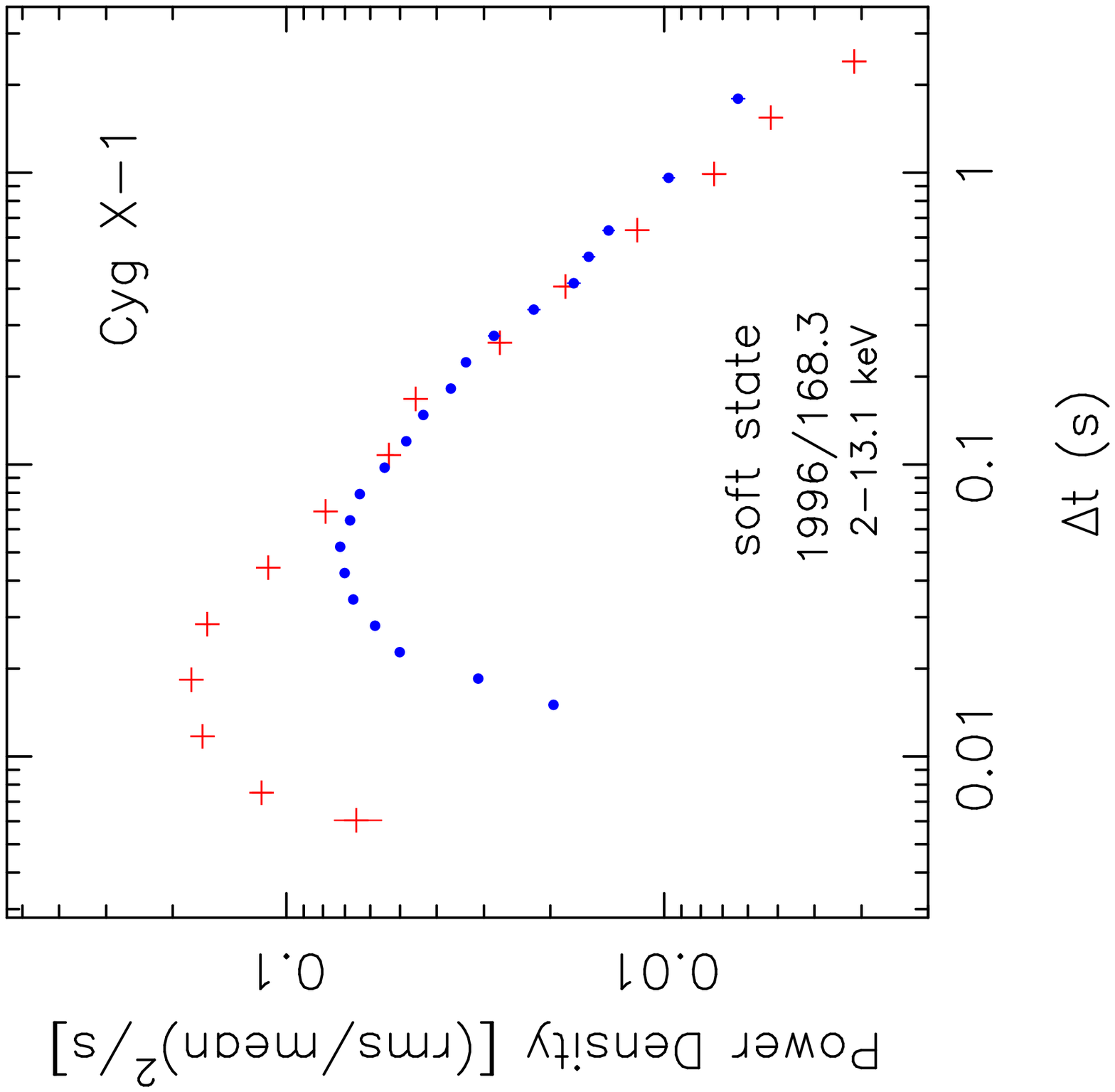}{20pt}{-90}{40}{25}{-50}{215}
\plotfiddle{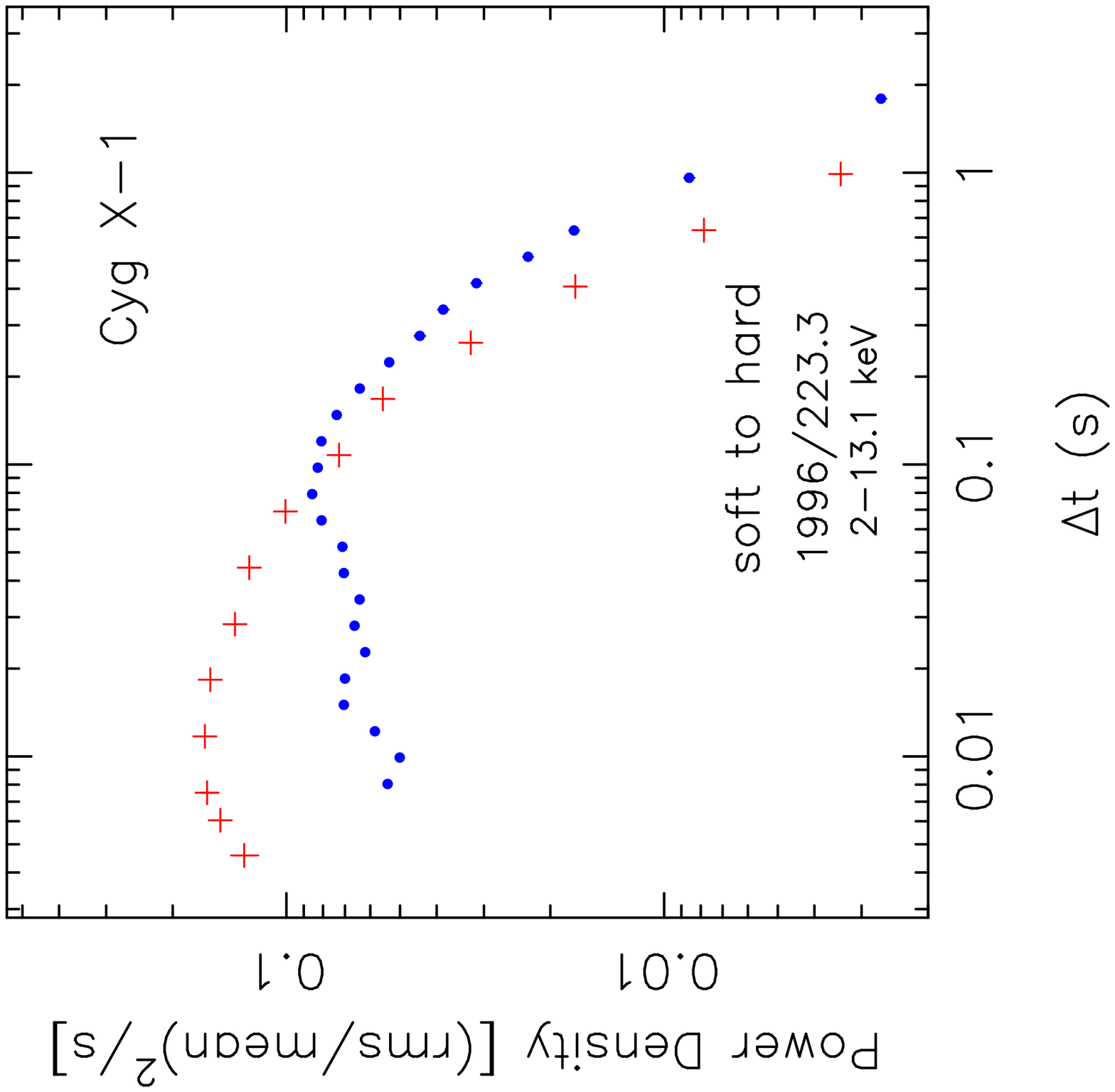}{20pt}{-90}{40}{25}{-260}{119}
\plotfiddle{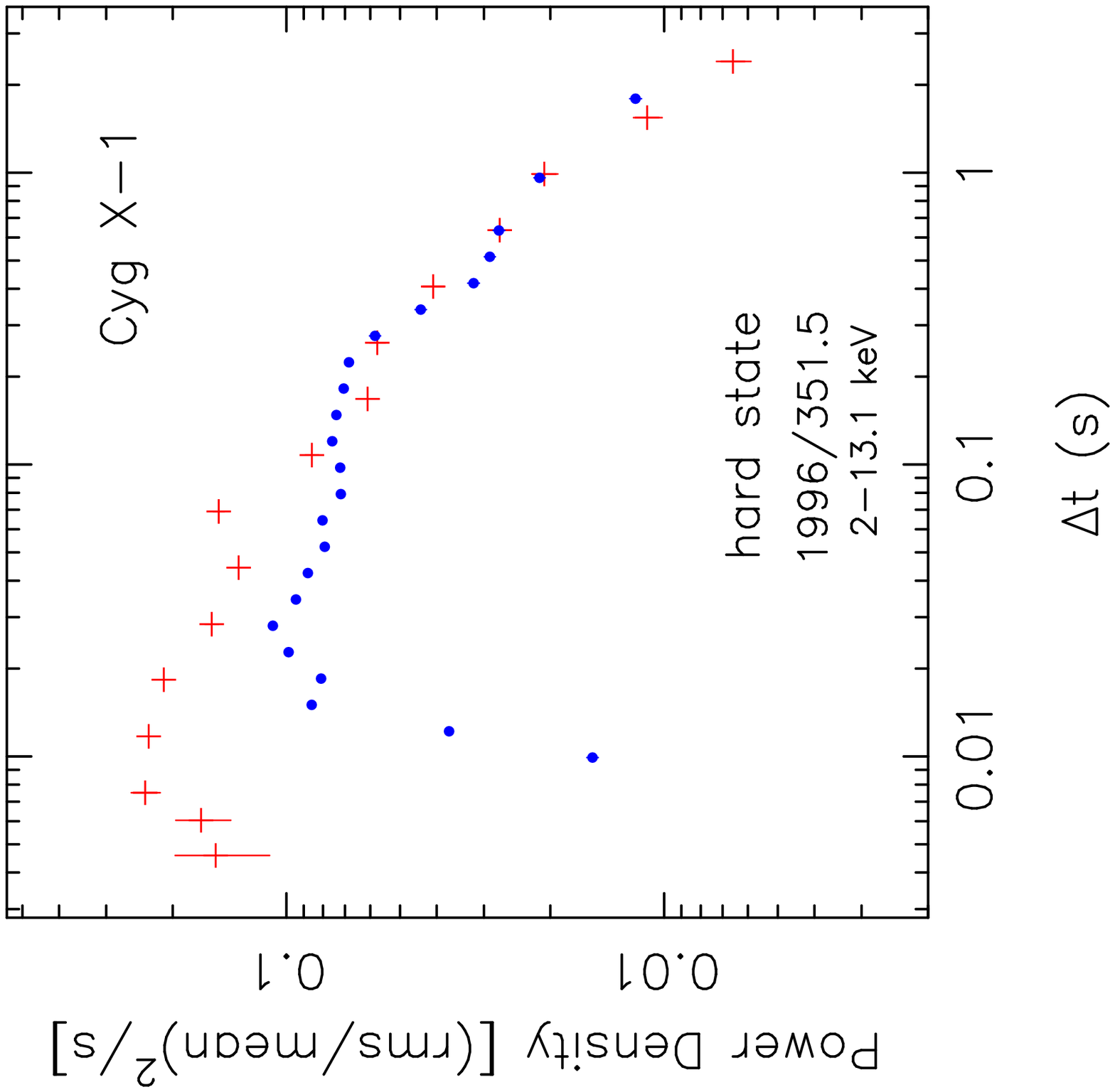}{20pt}{-90}{40}{25}{-50}{155}
\vspace{-1cm}
\caption{Power density vs. time scale of Cyg X-1.
{\it Plus}: spectrum in the time domain. {\it Dot}: Fourier
spectrum. 
\label{fig8}}
\end{figure}

\clearpage

\begin{figure}
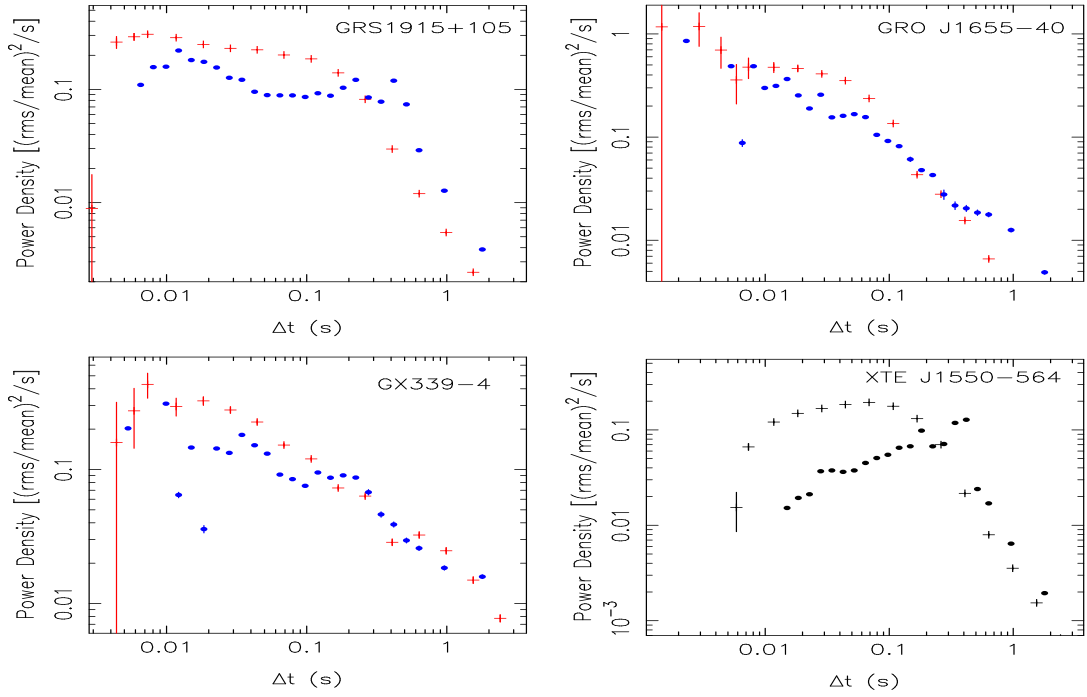

\vspace{3cm}
\epsscale{1.0}
\plotfiddle{f9a.eps}{20pt}{-90}{40}{25}{-260}{178}
\plotfiddle{f9b.eps}{20pt}{-90}{40}{25}{-50}{215}
\plotfiddle{f9c.eps}{20pt}{-90}{40}{25}{-260}{119}
\plotfiddle{f9d.eps}{20pt}{-90}{40}{25}{-50}{155}
\caption{Power density vs. time scale of 4 accreting black holes.
{\it Plus}: spectrum in the time domain. {\it Dot}: Fourier
spectrum. 
\label{fig9}}
\end{figure}


\begin{figure}
\vspace{2cm}
\epsscale{1.0}
\plotfiddle{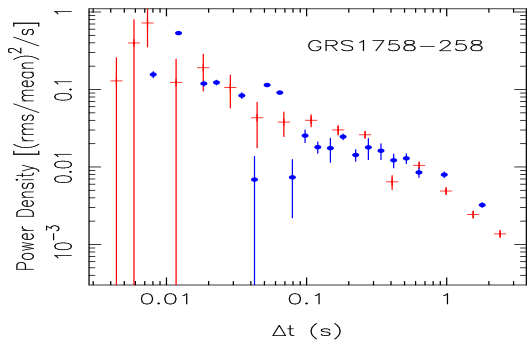}{20pt}{-90}{40}{25}{-260}{178}
\vspace{-10mm}
\caption{Power density vs. time scale of black hole binary GRS1758-258.
{\it Plus}: spectrum in the time domain. {\it Dot}: Fourier
spectrum. 
\label{fig10}}
\end{figure}

\end{document}